\documentclass[runningheads]{llncs}

\usepackage{graphicx}
\usepackage{amsmath,amssymb,amsfonts}
\usepackage{algorithmic}
\usepackage{cite}
\usepackage{xcolor}
\usepackage{subfig}
\usepackage[hyphens]{url}
\usepackage{hyperref}
\usepackage{listings}
\begin{document}
%\sloppy
%
\title{Using SGX for Meta-Transactions Support in Ethereum DApps}

%\subtitle{(work-in-progress paper)}
%\author{Emanuel Onica\inst{1,2} and Ciprian Amariei\inst{1}}

\author{Emanuel Onica and Ciprian Amariei}

%\institute{Alexandru Ioan Cuza University of Ia\c{s}i, Romania \and Eman Tech SRL, Romania \\ 
\institute{Alexandru Ioan Cuza University of Ia\c{s}i, Romania \\ % \and Eman Tech SRL, Romania 
\email{eonica@info.uaic.ro, ciprian.amariei@gmail.com} 
}
\maketitle              % typeset the header of the contribution
\vspace{-6mm}
\begin{abstract}
Decentralized applications (DApps) gained traction in the context of the blockchain technology. 
Ethereum is currently the public blockchain that backs the largest amount of the existing DApps. 
Onboarding new users to Ethereum DApps is a notoriously hard issue to solve. 
This is mainly caused by lack of cryptocurrency ownership, needed for transaction fees. 
Several meta-transaction patterns emerged for decoupling users from paying these fees. 
However, such solutions are mostly offered via off-chain, often paid relayer services and do not fully address the security issues present in the meta-transaction path. 
In this paper, we introduce a new meta-transaction architecture that makes use of the Intel Software Guard Extensions (SGX).  
Unlike other solutions, our approach would offer the possibility to deploy a fee-free Ethereum DApp on a web server that can directly relay meta-transactions to the Ethereum network while having essential security guarantees integrated by design.

\footnote{Author's version. The final authenticated publication is available online at https://doi.org/[to be completed].}
\end{abstract}
\vspace{-8mm}
\section{\uppercase{Introduction}}
\label{sec:introduction}

Blockchain networks created the context for developing new applications that leverage decentralized trust.
The role of nodes in a blockchain network is to maintain a replicated data structure, the main part of it being commonly referred as the ledger.
Nodes validate transactions sent by clients that change the replicated data.
Transaction blocks are formed and mutually agreed in a decentralized manner. 
Finally, confirmed blocks are appended to the ledger.

\begin{figure}[!ht]
\vspace{-3mm}
\centering
\hspace{0cm}
\begin{minipage}[b]{0.47\linewidth}	
  \centering
  \includegraphics[width=\linewidth]{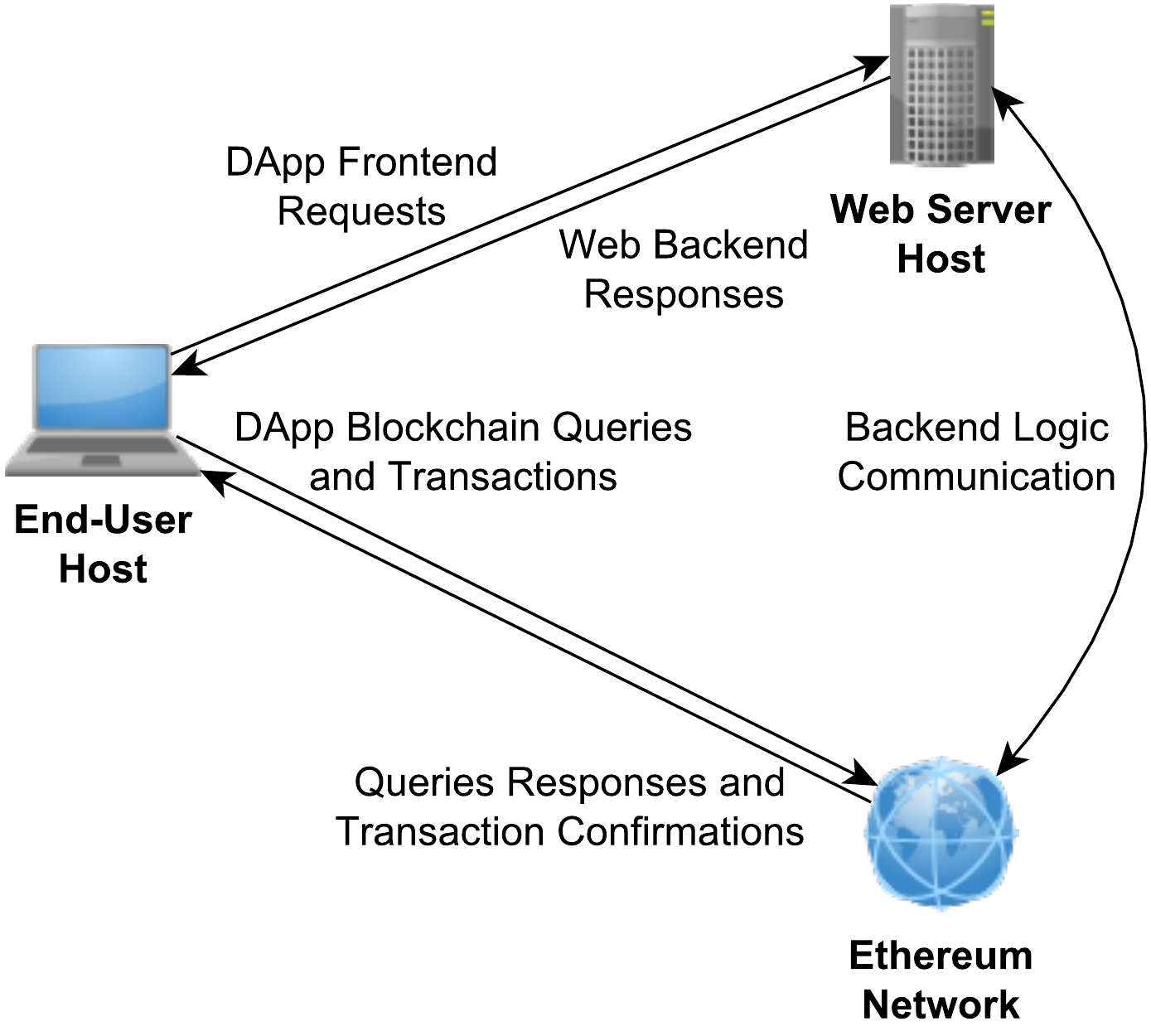}
  \vspace{-6mm}
  \caption{\label{fig.DApp} Typical flow for Ethereum DApp interaction.}
\end{minipage}
\hspace{1mm}
\begin{minipage}[b]{0.47\linewidth}	
 \centering
 \includegraphics[width=\linewidth]{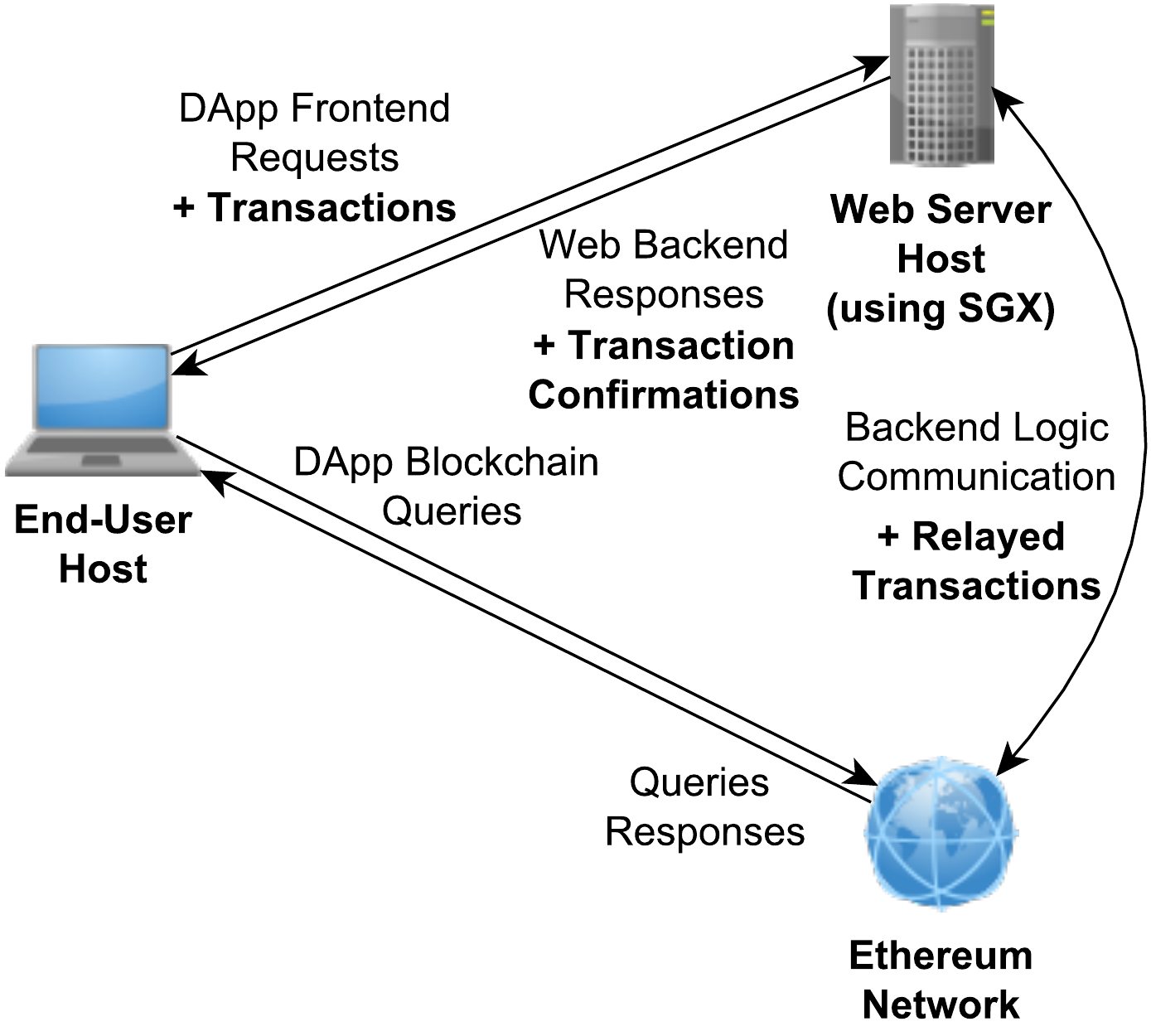}
 \vspace{-6mm}
 \caption{\label{fig.DAppTx} Interaction flow when relaying using SGX meta-transactions.\protect\footnotemark}
\end{minipage}
\vspace{-6mm}
\end{figure}

Newer blockchains provide support for smart contracts, small programs executed on the blockchain nodes.
Transactions can trigger functions operating over a contract state stored as part of the blockchain replicated data.
This significantly expanded the range of blockchain applications, from the fintech area to games and others, coined under the generic name of \emph{DApps}.
Ethereum~\cite{Ethereum} was the first platform to support smart contracts, and is still dominating the DApps market~\cite{DAppRadar,StateDApps}.
An Ethereum DApp is most often implemented as a web application deployed on a web server having part of its backend using smart contracts on the blockchain. 
The frontend can query the blockchain for information.
Also, actions performed by end-users can trigger transactions to smart contracts.
A simplified view of the DApp interaction flow is depicted in Figure~\ref{fig.DApp}.

Users onboarding is a known major issue in Ethereum DApps development~\cite{OnboardingBillion,GasSpectrum,UniLogin}. 
This stems from the requirements a user must fulfill for enabling DApp interaction with the blockchain backend. 
Ethereum transactions charge a fee. 
This fee is required to regulate the transaction processing load and as incentive for the network nodes, and must be paid by the transaction initiator.
This implies Ethereum cryptocurrency ownership by the user of the application.
Unfortunately this prevents DApps to target many users that might not even be familiar with the notion of cryptocurrency or simply are not willing to pay. 

~\emph{Meta-transactions}~\cite{Griffith18,Rush17} emerged as a solution for users onboarding.
In essence, this implies wrapping end-user transaction data in transactions paid by a different entity, which can be either the DApp owner or another sponsor. 
Although the concept seems simple, the implementation and deployment are not.
Some important issues arise when integrating meta-transaction relaying with a DApp. 
The funds paying for transactions must be secured, as well as the integrity of the end-user transaction data. 
The few maintained solutions are typically offered as third party relayer services~\cite{GSN,ITX,Biconomy}.  
These address transaction integrity but disregard the protection of funds allocated for paying transactions.
Some also charge a relayer fee or require consistent changes in the DApp architecture. 
This complicates the task of a developer in finding a suitable DApp design.

In this paper we introduce the \emph{SGX meta-transaction} architecture, intended to facilitate secure meta-transaction relaying integration for DApp developers. 
Our purpose is to permit meta-transaction wrapping to be handled securely by the DApp host, which will act as transaction relayer.
For this, we use a trusted execution environment (TEE), namely the Intel Software Guard Extensions (SGX)~\cite{SGX}. 
This changes the transaction path as depicted in Figure~\ref{fig.DAppTx}.

\footnotetext{Simple queries not changing the smart contract state are considered free in Ethereum, otherwise these should follow the same path as transactions.}

Our paper is structured as follows.
In Section~\ref{sec:background} we present some background on Ethereum DApps and the context of meta-transactions.
In Section~\ref{sec:design} we introduce our architecture and an initial proof-of-concept implementation. 
We discuss some extensions in Section~\ref{sec:directions}.
Finally, we conclude in Section~\ref{sec:conclusion}.

\section{\uppercase{Background}}
\label{sec:background}

Users interacting with Ethereum DApps can trigger transactions, such as cryptocurrency transfers or calling functions in smart contracts that change the blockchain data. 
The latter is the more general case and our focus. 
Transactions come at a cost quantified in~\emph{gas} units. 
This cost increases with the complexity of operations executed in the smart contract.
The user must pay a transaction fee equal to the cost in gas multiplied with a price per gas unit set in the Ethereum cryptocurrency.
This price per gas unit is composed of a variable base network fee to which a priority fee can be added to speed up transaction processing. 

Two types of accounts are defined in Ethereum: externally owned accounts (EOAs) and contract accounts. 
Any account is identified by an address and has a balance in the Ethereum currency.
Transactions can be submitted by EOAs, essentially user accounts controlled by private keys used to sign the transactions.
The fees of verified transactions are deducted from the EOA balance.
The main part included in a transaction message is either or both of a data payload encoding a smart contract function call and a currency value to be transferred.
Other transaction fields include an incremental nonce bound to the EOA, a gas limit, the maximum gas price, the recipient address and the EOA's signature. 

We consider DApps where the interaction does not imply a payment and users can have a zero balance in Ethereum currency. 
In such cases, a~\emph{meta-transaction} would wrap the original end-user's transaction data, and must be signed and paid by an EOA address capable of covering the transaction fees.
The DApp developer is faced with the challenge of implementing a signature delegation pattern to such an EOA address, providing appropriate trust guarantees. 

Deployed solutions typically require DApps to use off-chain relayer services~\cite{GSN,ITX,Biconomy}.
Integrating a third party service into the transaction path comes with an inherent risk to the transaction integrity.
Therefore, these solutions focus on ensuring that the service itself cannot tamper the original data when wrapping it into a meta-transaction. 
Provided APIs require the user's EOA signature to be present in their sent data and to adapt the smart contracts backend of the DApp to verify that. 
However, this does not protect the private key used for signing the meta-transaction itself.
The relayer service must be provided with funds for paying the meta-transaction.  
This makes critical storing securely the relayer's signing key. 
If an attacker gains access to this key it can drain the relayer funds, by simply signing transactions transferring the relayer's balance to the attacker. 

In a normal transaction scenario, keeping the signing key safe is solely the responsibility of the end-user who operates with her own funds. 
In the relayed meta-transaction scenario this guarantee should be provided by the relayer.
Unfortunately, none of the relayer implementations we are aware of offers details on how it secures the meta-transaction signing key.
Some relayer providers do not even specify whether they host their service on their private infrastructure or on a public cloud, case proven vulnerable to sensitive data leaks~\cite{Ristenpart2009,Zhang2014,Varadarajan2015}.

We propose a meta-transaction architecture that does not depend on an external relayer and overcomes the security issues above.
This simplifies integrating meta-transaction support in a DApp and saves fees charged by external relayers. 

\section{\uppercase{Basic Solution Design}}
\label{sec:design}

The purpose of our design is to provide easy integration for safe meta-transaction support with the DApp backend implementation and to use the DApp host as a secure relayer.
Eliminating a third party relayer service from the transaction path automatically eliminates the concern of this party tampering with the transactions.
However, we consider the web server host where the DApp is deployed untrusted with respect to preserving the confidentiality of sensitive information.
The main threat we tackle is an attack trying to leak private credentials from this host, such as the key used in signing the meta-transactions.

To prevent private key leakage we employ the use of Intel SGX, a widely available TEE solution.
Its core abstraction is an~\emph{enclave}, which isolates sensitive code execution within an encrypted memory region.
An enclave implementation can provide a set of functions -~\emph{ECalls}, to be called from untrusted code outside the enclave for executing code in secure isolation within the enclave.
Another set of functions, the~\emph{OCalls} are used when the code inside the enclave initiates calls to untrusted code. 
The definition of ECalls and OCalls forms the interface of the enclave.
An enclave can be remotely attested in order to verify the integrity of the enclave code and if this is executed on a genuine SGX capable processor.
The remote attestation can also be used to establish a shared secret base for encrypted communication between the enclave and the party requesting the attestation.

We use an SGX enclave integrated with the DApp for the sensitive operations in the transaction flow.
Once the DApp is deployed, the DApp owner must execute an enclave initialization protocol.
This protocol establishes a set of~\emph{master credentials}, namely an Ethereum account address and the corresponding signing key, to be used within the enclave. 
These credentials are randomly generated in the enclave and can be sent to the DApp owner via a secure channel established as part of the attestation. 
The DApp owner will use the master account address to transfer funds for covering the meta-transaction fees.
The master signing key must be safely stored by the DApp owner and is not used in normal operation outside enclave space.
This key is required to be sent to the DApp owner only to maintain control over the funds in case of enclave failure.

\begin{figure*}[!ht]
\vspace{-1mm}
   \centering
	\includegraphics[width=0.85\textwidth]{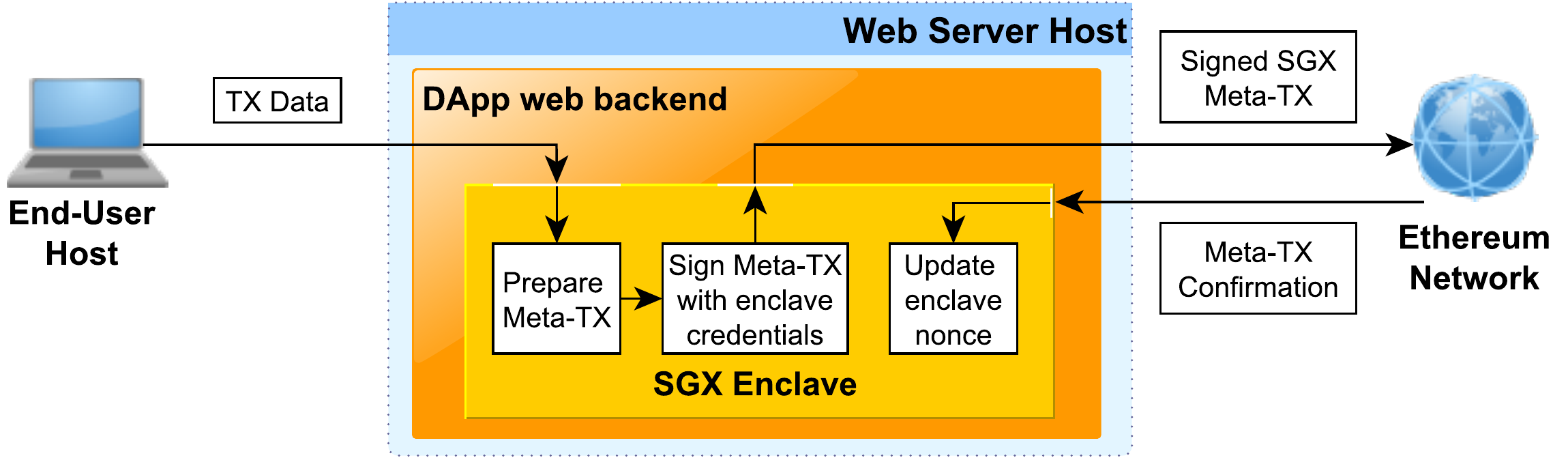}
	\caption{High level overview of the SGX meta-transaction flow.}
	\label{fig.TXtraffic}
\vspace{-5mm}
\end{figure*}

After this initialization the enclave is ready to operate on transaction data sent by a user. 
We define a~\emph{SGX meta-transaction} as a meta-transaction prepared and signed within the secure enclave space.
A simplified overview of the enclave integration within the transaction flow is presented in Figure~\ref{fig.TXtraffic}.
The transaction data contains the serialized encoding of the smart contract function call and the contract address. 
This is received at the DApp web backend and passed via an ECall to the enclave. 
Additional information necessary to form an Ethereum transaction such as gas related parameters can be passed with the transaction data or established in the enclave space. 
The SGX meta-transaction is prepared within the enclave using an encoding required by Ethereum, wrapping the data and the rest of fields including a sequentially increasing nonce.
This nonce is associated to the enclave's master Ethereum address and is required for transaction ordering.
Finally, the enclave code signs the SGX meta-transaction using the master signing key and passes it to the web backend through an OCall.

Following the above steps, the web backend code of the DApp can relay the signed SGX meta-transaction to the Ethereum blockchain.
The enclave maintains a trusted keystore, secured using the sealing key - a hardware key unique per CPU accessible only in the enclave. 
The keystore is loaded in the enclave memory when needed and can be stored encrypted on disk. 
The structure of this keystore can be adapted to fit the needs of the DApp.
In its simplest form it holds the set of master credentials. 
Once the web backend receives the transaction confirmation an ECall will trigger the nonce increment in the keystore.

We have implemented a proof-of-concept of the above design wrapping the SGX enclave within a native Node.js module~\cite{Node}. 
This module allows the DApp backend to trigger the necessary meta-transaction flow operations within the enclave. 
Most Ethereum DApp implementations use JavaScript libraries~\cite{Web3, Ethers} for interacting with the blockchain network.
Therefore, providing our solution as a Node.js module makes seamless the integration with most DApps. 
We performed a functionality test of our transaction flow on a mockup DApp where the user can change a value in a smart contract deployed on the Ethereum Ropsten test network. 
Our SGX meta-transaction constructed  within the enclave was successfully validated by the network.\footnote{The record of the first SGX meta-transaction relayed via our implementation is available at the following address: \url{https://ropsten.etherscan.io/tx/0xdcb13cdaaf847ddce26307988ac4938c9037e03b747276f46b222df2a42d302b}}
We tested the implementation on a SGX capable machine equipped with an Intel i7-7700 CPU running Ubuntu 18.04.5 LTS.
The measured time overhead for preparing the signed SGX meta-transaction was in the range of 3ms including logging, orders of magnitude smaller than the average confirmation time of an Ethereum transactions block at almost 14s.

\vspace{-3mm}
\section{\uppercase{Discussion and Open Directions}}
\label{sec:directions}
\vspace{-1mm}

The description in the previous section is limited to the bare necessities in the transaction flow.
In the following we examine some of the extensions we consider. 

A more complex structure of the keystore could include multiple Ethereum credentials generated for signing meta-transactions.
This scenario could fit allocating separate funds for different users or attempts to scale the transaction flow.
Exporting multiple addresses and safely storing their private keys would be, however, prone to increased security risks for the DApp owner.
Therefore, for such a scenario we consider keeping these keys confined in the enclave space.
The master account address would act as a central deposit for funding the meta-transactions signed by each of the secondary accounts.
This would be done by periodical value transactions sent to these internal accounts and will obviously add an extra cost. 
However, a simple value transaction has the smallest cost in Ethereum and tuning the periodicity of funding can minimize the overhead.

A particular case is of DApps where the Ethereum identity of a user must be preserved in the transactions: 
DApps using tokens, either fungible, essentially virtual coins built over Ethereum, or NFTs. 
The approach in other solutions~\cite{GSN,Griffith18,Griffith18Native,ITX} is to include a signature using user's own Ethereum credentials in the meta-transaction and adapt the smart contract logic to verify it. 
Our design in Section~\ref{sec:design} can easily accommodate such changes in the carried transaction data.

EIP-2771~\cite{eip2771} proposes a contract level protocol for validating data signed with user's Ethereum credentials in meta-transactions.
While we can integrate our solution also with this architecture, we note that its main scope is to guarantee integrity against a relayer controlled by an untrusted third party. 
In our design the DApp owner controls the relayer.
Nevertheless, we could consider a possible integrity attack escalation over the web server. 
This can be mitigated by a TLS channel terminated within the enclave over which the end-user will send the transaction data.
This guarantees the integrity up to the enclave on the relaying host.
Further, the SGX meta-transaction is securely signed in the enclave, therefore it cannot be altered until verified in the blockchain network.
We have considered various TLS implementations in conjunction with SGX for such an extension~\cite{IntelSGXSSL,wolfSSL,talosSGX,Gramine}. 
Some provide performance advantages, while others seem to be easier to integrate with our web oriented architecture.
For brevity we leave further technical details for a future extended report of our work.

Finally, an aspect to consider is the solution deployment. 
An attractive option would be to deploy the DApp over a public cloud platform. 
Currently the support for SGX offered in virtualized environments comes with a performance impact as discussed in~\cite{SGXVMS}.
Further analysis is required, but we believe the transaction confirmation time plus the network latency would still overshadow the additional penalties inflicted by the virtualization.
\vspace{-3mm}

\section{\uppercase{Conclusion}}
\label{sec:conclusion}
\vspace{-1mm}

We introduced in this paper a new architecture for relaying Ethereum meta-transactions.
Unlike external, sometimes paid services, our solution takes a different approach aiming for a secure integration of meta-transaction relaying support directly within the DApp. 
Our design introduces the SGX meta-transaction prepared and signed within a secure enclave space.
This provides independence to a DApp developer, it relaxes integrity concerns by not needing to trust an extra third party 
and offers solid guarantees on preventing leaks that could lead to losing funds allocated for paying the meta-transaction fees. 

We emphasize that our proposed architecture is a work-in-progress. 
We briefly discussed multiple extensions we consider. 
We believe that our proof-of-concept integrating SGX meta-transactions via a Node.js module already shows the practicality of our design and promising potential for use within DApps.

\bibliographystyle{splncs04}
\bibliography{references}

\end{document}